\documentclass{ifacconf}

\usepackage{graphicx} 
\usepackage{natbib} 
\usepackage{float}
\usepackage{caption}
\usepackage{subcaption}
\usepackage{textcomp}
\usepackage{xcolor}
\usepackage{amssymb,amsfonts}
\usepackage{algorithm} 
\usepackage{algpseudocode}
\usepackage{latexsym}
\usepackage{amsmath}
\usepackage{physics}

\DeclareMathOperator*{\argmax}{arg\,max}

\begin{document}
\begin{frontmatter}

\title{Deep Reinforcement Learning Based Controller for Active Heave Compensation}

\author[First]{Shrenik Zinage} 
\author[First]{Abhilash Somayajula} 

\address[First]{Department of Ocean Engineering, Indian Institute of Technology Madras (IITM), Chennai, India, 600036 (e-mail: shrenikvz@gmail.com, abhilash@iitm.ac.in)}

\begin{abstract}                
Heave compensation is an essential part in various offshore operations. It is used in various applications, which include on-loading or off-loading systems, offshore drilling, landing helicopter on oscillating structures, and deploying and retrieving manned submersibles. In this paper, a reinforcement learning (RL) based controller is proposed for active heave compensation using a deep deterministic policy gradient (DDPG) algorithm. A DDPG algorithm  which is a model-free, online reinforcement learning method, is adopted to capture the experience of the agent during the training trials. The simulation results demonstrate up to $10$ $\%$ better heave compensation performance of RL controller as compared to a tuned Proportional-Derivative Control. The performance of the proposed method is compared with respect to heave compensation, offset tracking, disturbance rejection, and noise attenuation.
\end{abstract}

\begin{keyword}
deep reinforcement learning, learning-based control, winch control, artificial intelligence, active heave compensation
\end{keyword}

\end{frontmatter}

\section{Introduction}

With the increase in the number of ocean explorations and a huge demand for various marine resources, heave compensation has become a vital part of various maritime operations. In heave compensation, the primary objective is to decouple the motion of a payload connected to the ship from the ship's vertical (heave) motion. Heave compensation methods can be broadly classified into two main categories: passive heave compensation (PHC) and active heave compensation (AHC). The PHC is an open-loop system that is designed to partially decouple the payload from the vessel. The compensation performance for PHC is generally observed to be less than 80 \% (\cite{hatleskog2007passive}). AHC relies on closed-loop control system architecture and provides the payload displacement as a continuous feedback to the controller so that an improved compensation performance is achieved. 

Over the years, various classical control techniques have been analysed and compared to keep a suspended payload regulated when the vessel is undergoing dynamic vertical motion in the ocean (\cite{zinage2020comparative,li2019adrc,woodacre2018hydraulic,do2008nonlinear}). However, not much research has looked at using reinforcement learning (RL) control techniques, which is one of the methods gaining popularity in marine applications (\cite{woo2019deep,martinsen2018straight,zhao2019colregs}).

Lately, model-free deep reinforcement learning has
made significant progress in solving a variety of complex tasks. The first successful application of this technique to learn the control policy was through a deep Q-network (DQN) for playing Atari games (\cite{mnih2013playing,silver2017mastering}), which integrates the Q learning and deep neural network. However, DQN can only be used to solve problems that have discrete action space. Since many control tasks in the real world have continuous action space, several advanced reinforcement learning algorithms aiming to solve continuous control problems have also been developed (\cite{lillicrap2015continuous,mnih2016asynchronous,levine2016end}).

In this paper, a deep deterministic policy gradient (DDPG) (\cite{lillicrap2015continuous}) algorithm that is based on an actor-critic framework is used. This method has an advantage over the DQN approach that it can deal with a continuous action space. Apart from that, this algorithm inherits conventional approaches of RL such as actor-critic (\cite{sutton1999policy}), and policy gradient (\cite{konda2000actor}).

\emph{Main Contributions}: This paper introduces a RL based control methodology using the DDPG algorithm for active heave compensation. The efficacy and the potential of using this technique for real-life applications are assessed in a simulated environment. To the best of our knowledge, this is the first time that RL is being investigated as a tool for active heave compensation.  

\emph{Structure of the paper}: The organisation of the paper is as follows: Section 2 discusses the modeling of the ship's motion. The dynamic model of the winch is described in Section 3. Section 4 describes the theoretical background and problem formulation of RL based controller. The simulation results are presented in Section 5 and the conclusions are presented in Section 6.

\section{Modeling of Ship Motion}

In this paper, the KRISO container ship (KCS) is chosen for calculating the dynamic motion in the ocean. 
\begin{table}[htbp]
\begin{center}
\captionsetup{width=8.5cm}
\caption{KCS Particulars}
\label{tb:kcs_particulars}
\resizebox{8.5 cm}{!}{\begin{tabular}{lr}
\textbf{Particulars} & \textbf{Value}\\\hline
Length between perpendiculars $L_{pp}$ & 230 $m$\\
Length waterline $L_{WL}$ & 232.5 $m$\\
Breadth $B$ & 32 $m$ \\
Depth $D$ & 19 $m$ \\
Draft $T$ & 10.8 $m$ \\
Displacement & 52030 $m^3$ \\
Block Coefficient $C_{B}$ & 0.65 \\
LCB from midship (fwd +) & -3.404 $m$ \\
LCB from AP (fwd +) & 111.596 $m$ \\
VCG from WL & 3.551 $m$ \\
VCG from keel & 14.351 $m$ \\
GM & 0.6 $m$ \\
Design Forward Speed $U$ &  24 $knots$ \\
Analysis Speed (In this study) & 0 $knots$\\
Froude Number $F_n$ & 0.26 \\
Roll Radius of gyration about CG $k_{xx}$ & 12.88 $m$\\
Pitch/Yaw radius of gyration about CG $k_{yy}/k_{zz}$ & 57.5 $m$\\\hline
\end{tabular}}
\end{center}
\end{table}
Table~\ref{tb:kcs_particulars} shows the particulars of an KCS ship. The Pierson Moskowitz (PM) wave elevation spectrum corresponding to a significant wave height $H_s$ and peak period $T_p$ for a range of frequencies is defined by 
\begin{equation}
    S(\omega) = \frac{0.31}{2\pi}T_pH_s^2\left(\frac{\omega T_p}{2\pi}\right)^{-5}exp\left(\frac{-5}{4}\left(\frac{\omega T_p}{2\pi}\right)^{-4}\right)\label{eq:1}
\end{equation}
An irregular wave elevation time history is generated from the above spectrum. The wave elevation is expressed as a sum of $N$ sinusoidal components as shown below
\begin{equation}
    \eta(t) = \sum_{i=1}^{N} A_n cos(\omega_{n}t + \phi_{n})\label{eq:2}
\end{equation}
In this study, $N$ is taken as $100001$ to simulate a $10000$ s time history with a sampling time of $0.1$ s. 
For a simulation of duration $T$ seconds, the frequency increment is given by $\Delta \omega = 2\pi/T$. At each discrete frequency $\omega_{n} = n \Delta \omega$, the amplitude of the $n^{th}$ wave component is given by
\begin{equation}
    A_n = \sqrt{2S(\omega_n)\Delta \omega}\label{eq:3}
\end{equation}
The phase $\phi_n$ of each wave component is sampled from a uniform distribution between $-\pi$ and $\pi$ (\cite{Somayajula2017}). In order to avoid repeating of the generated signal the frequencies are randomised as shown below
\begin{equation}
    (\omega_n)_{new} = (\omega_n)_{old} + \Delta \omega  X
    \label{eq:4}
\end{equation}
where $X$ is a random variable following an uniform distribution between $-0.5$ and $0.5$. 

The response amplitude operator (RAO) of the KCS container ship is obtained for the heave, roll and pitch modes using MDLHydroD developed by \cite{guha2016development} which is a frequency domain 3D panel method based tool for analysis of wave structure interaction. 

Once the RAO is obtained, the response spectrum of the roll, pitch and heave is calculated as shown below
\begin{equation}
    S_\text{response}(\omega) = \mid H(\omega) \mid ^2S(\omega)
    \label{eq:5}
\end{equation}
where $S_\text{response}$ is the response spectrum, $H(\omega)$ is the RAO and $S(\omega)$ is the wave spectrum. Now the input wave elevation time history is decomposed into its frequency components by taking a fast Fourier transform (FFT). The amplitude of the response at a frequency $\omega_k = (k-1)\Delta \omega$ is then obtained by taking the product of RAO at that frequency and the FFT of input wave elevation time history at the same frequency.
\begin{figure}[ht]
\begin{center}
\includegraphics[width=8cm]{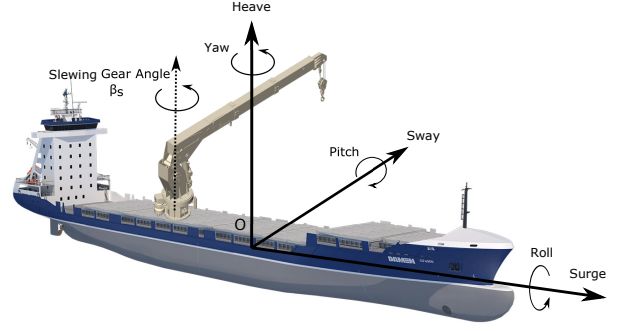}
\caption{Ship with Installed Crane } 
\label{fig:crane}
\end{center}
\end{figure}
Finally, inverse fast Fourier transform (IFFT) is used to convert these three degrees of freedom back into time domain.

In this study, the origin of the ship fixed coordinate system is assumed to be at the intersection of the waterline, centerline, and midship. Assuming that the crane is placed at $(x_\text{crane},y_\text{crane})$ in the vessel's body fixed coordinate system with a slewing gear angle $\beta_\text{s}$ and the wave incident angle of $\beta$, the net heave response time history of the winch placed on board the KCS ship is calculated in terms of the combined roll, heave, and pitch motion caused due to the wave excitation. Fig. \ref{fig:crane} shows a schematic diagram of the ship with a crane installed on it. Assuming small amplitude motions consistent with linear hydrodynamic theory, the net heave motion time history is given by
\begin{equation} 
\begin{array}{ll}
 z_\text{winch} = & \eta_\text{3}(\beta) + (y_\text{crane}+ l_\text{crane}sin(\beta_\text{s}))\eta_\text{4}(\beta) \\[0.2 cm] 
 & - (x_\text{crane} + l_\text{crane}cos(\beta_\text{s}))\eta_\text{5}(\beta)   
 \label{eq:6}
\end{array}
\end{equation}
where $\eta_\text{3}(\beta)$, $\eta_\text{4}(\beta)$, and $\eta_\text{5}(\beta)$ are the heave, roll and pitch time histories respectively, which depend on the incident wave angle $\beta$. In this study, the coordinate of the crane with respect to vessel's body frame is assumed to be at ($-1.5$ m, $2$ m) with a slewing gear angle of $30$ degrees and the horizontal extent of the crane ($l_{crane}$) is assumed to
\begin{figure}[ht] 
    \centering
  \subfloat[]{%
      \includegraphics[width=0.49\linewidth]{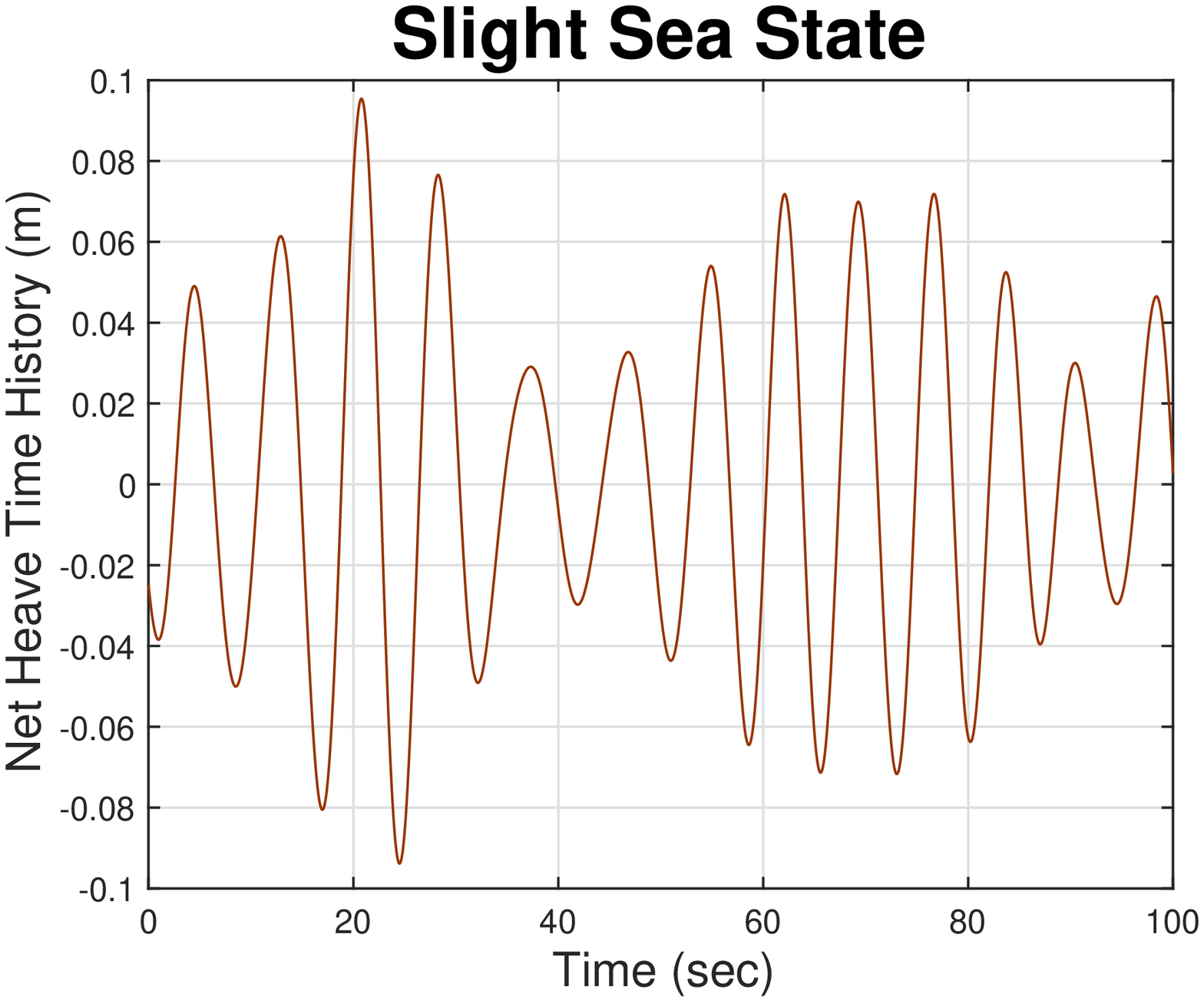}}
    \hfill
  \subfloat[]{%
        \includegraphics[width=0.49\linewidth]{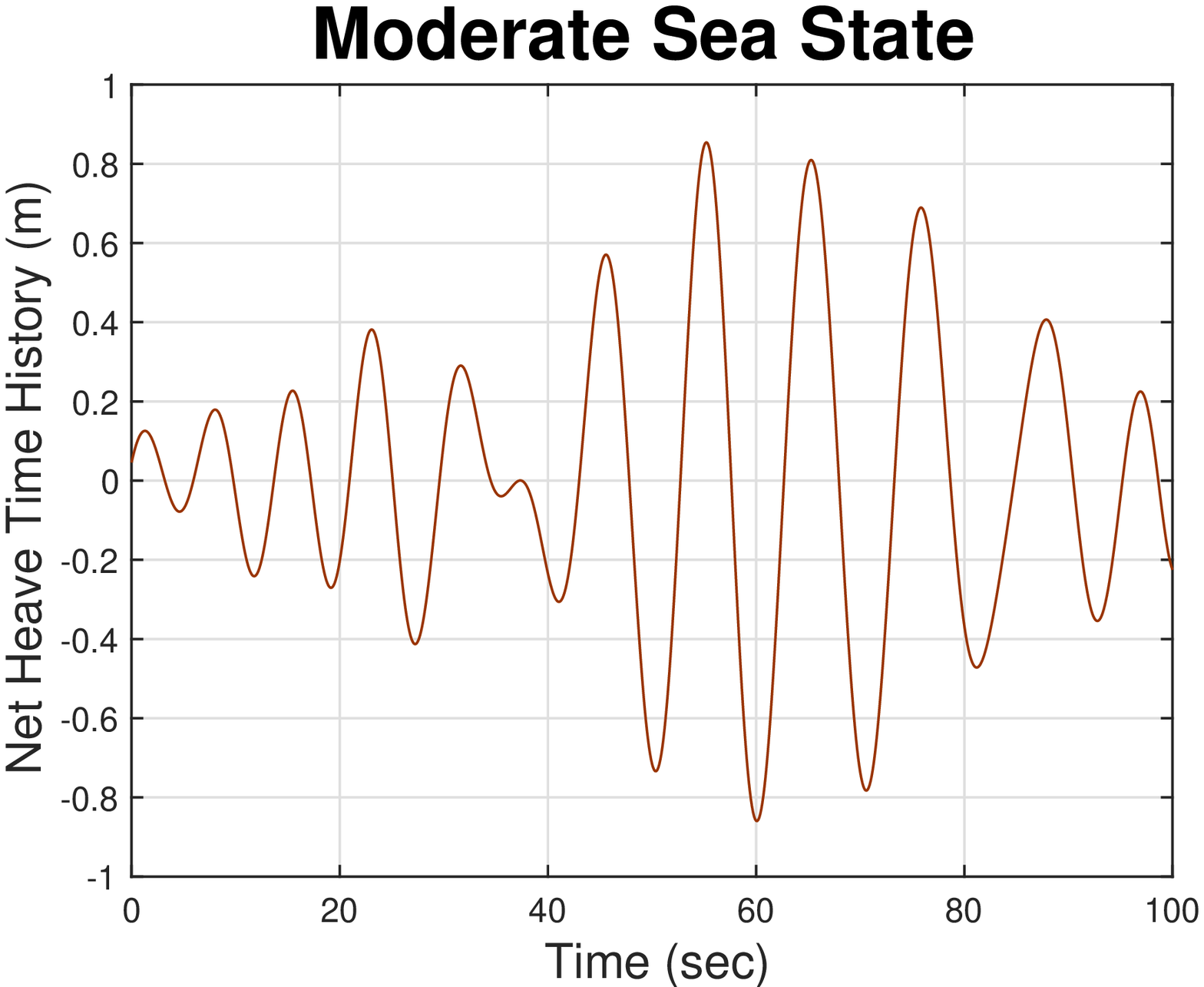}}
    \\
  \subfloat[]{%
        \includegraphics[width=0.49\linewidth]{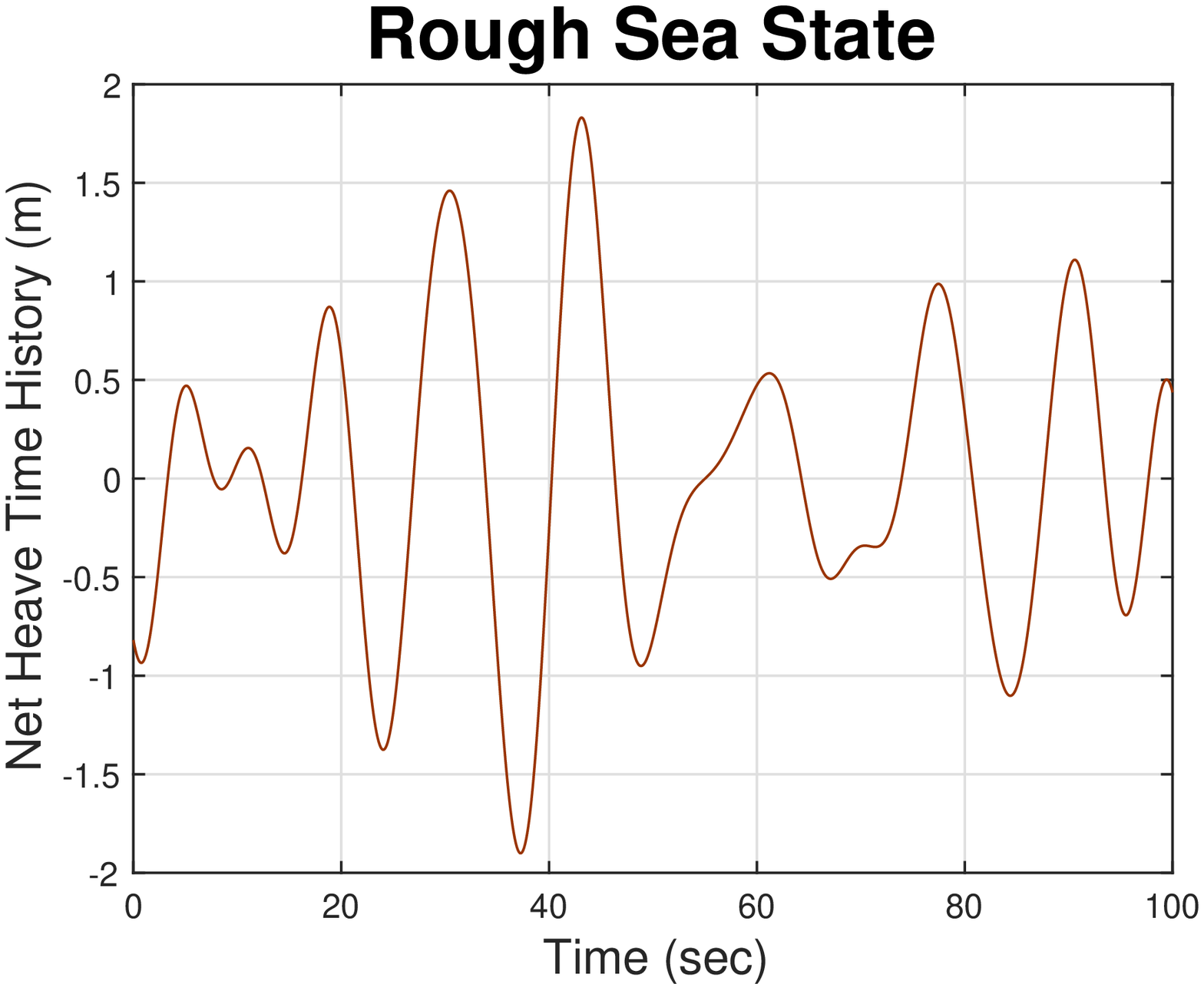}}
    \hfill
  \subfloat[]{%
        \includegraphics[width=0.49\linewidth]{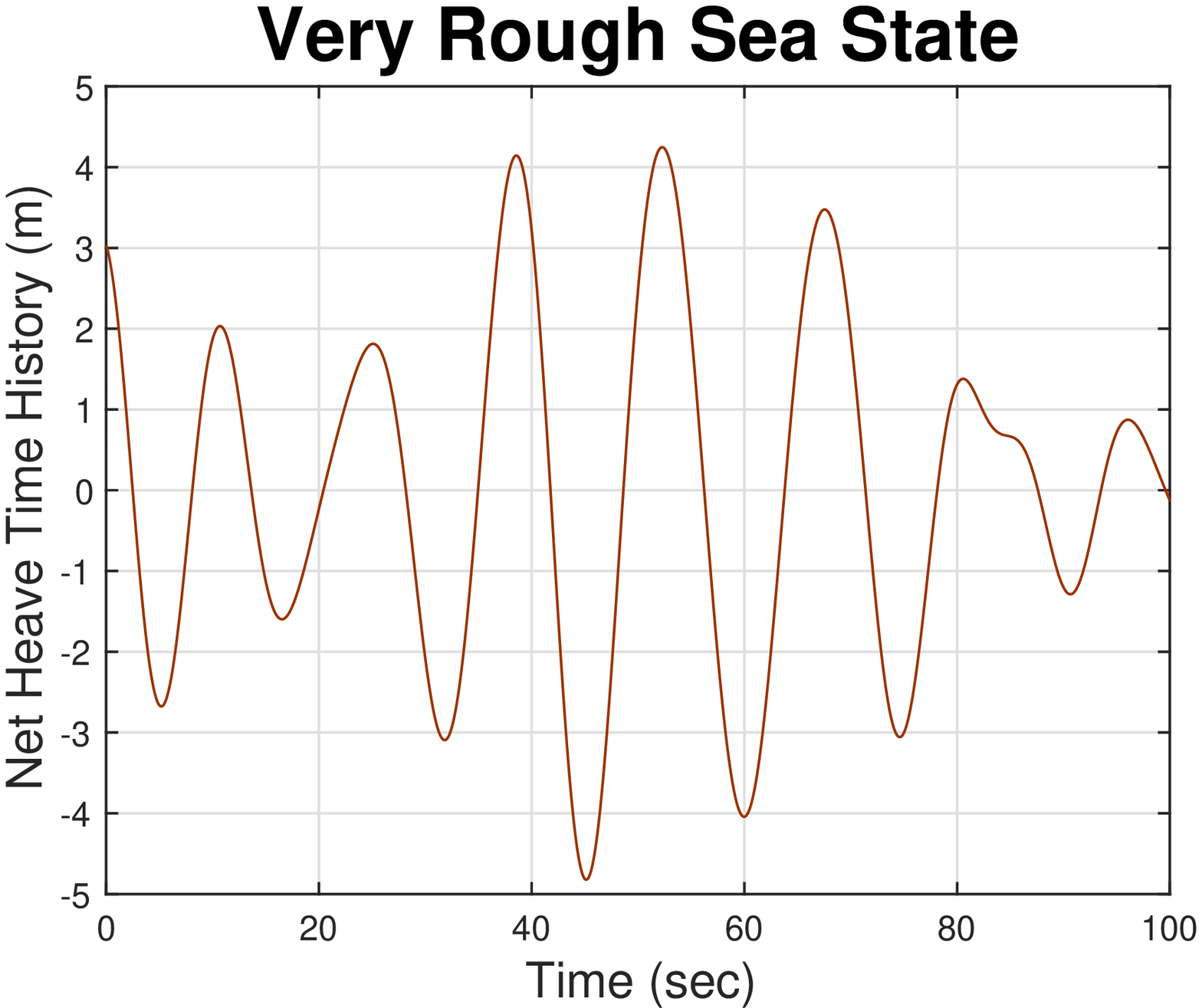}}
    \caption{Net heave time history generated from a PM spectra having (a) $H_s$ = $1.5$ m, $T_p$ = $6$ s (b) $H_s$ = $4$ m, $T_p$ = $9$ s (c) $H_s$ = $6$ m, $T_p$ = $12$ s (d) $H_s$ = $8.5$ m, $ T_p$ = $14$ s }
 \label{fig:heave_timehistory}
\end{figure}
be $3$ m. The plot of a $100$ second snip of the net heave motion time history in 4 different sea states when the waves are incident at an angle of $135$ degrees are shown in Fig. \ref{fig:heave_timehistory}.

\section{State Space Model}

\begin{figure}[ht]
\begin{center}
\includegraphics[width=8.4cm]{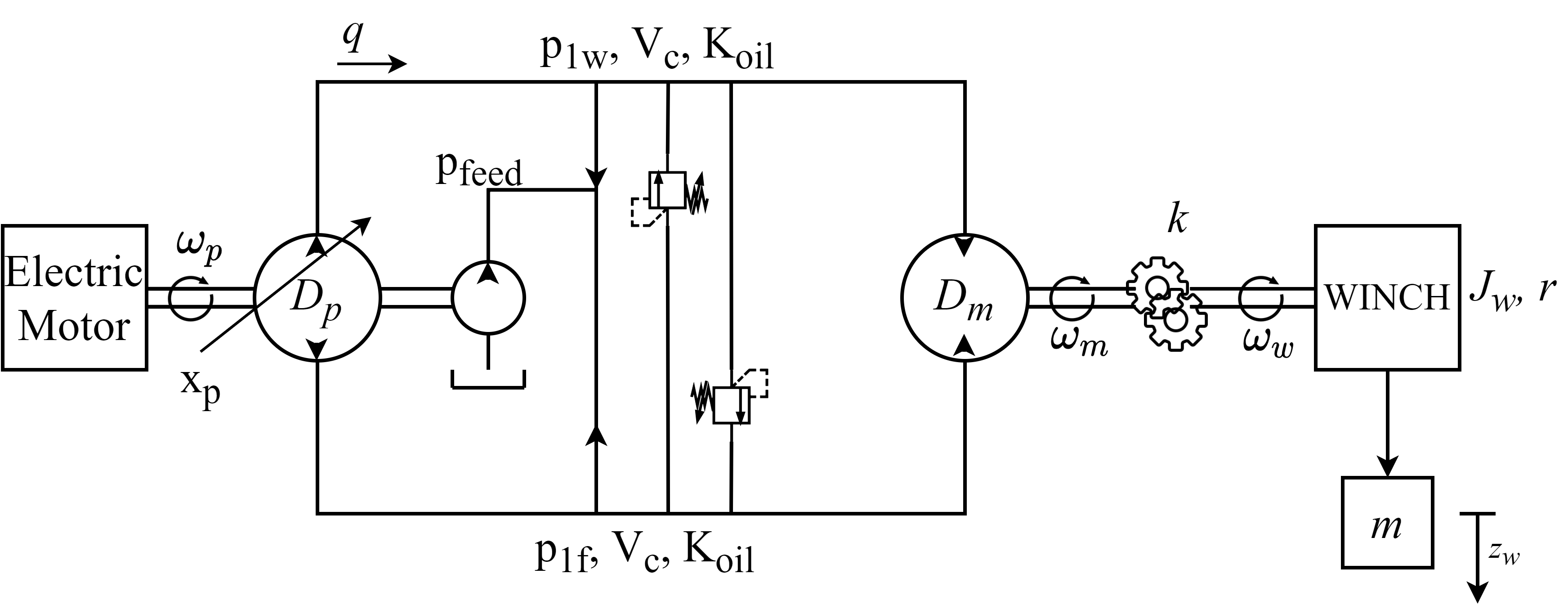}
\caption{Schematic diagram of hydraulic driven winch}
\label{fig:hydrualic}
\end{center}
\end{figure}

In this section, the state space model for hydraulic drive of the winch is described (\cite{richter2017experimental}). A schematic representation of the hydraulic drive is shown in Fig. \ref{fig:hydrualic}.

The state space model of the winch is given by
\begin{equation}
\begin{split}
    \dot{x} = & Ax + Bu_p + [0\quad 0\quad d\quad 0]^{T} \\[0.2 cm]
    y &= z_w = Cx, \quad  x(0) = x_0
    \label{eq:7}
\end{split}
\end{equation}
with the state
\begin{equation} 
   x = [x_p\quad \Delta{p}\quad \dot{z}_w\quad z_w]^{T}
   \label{eq:8}
\end{equation}
where $x_p$, $\Delta{p}$, $\dot{z}_w$, $z_w$ are normalized swash angle, change in pressure, velocity of the reeled rope, and length of the reeled rope respectively. The input to this plant is $u_p$ that translates to the normalized swash angle through a first order system as shown in \eqref{eq:9}.
\begin{equation}
    \dot{x}_p = \frac{-1}{T_w} (x_{p} - u_{p})
    \label{eq:9}
\end{equation}
A saturation of normalized swash angle is assumed beyond $x_p = \pm 1 $. The system matrix is given by 
\begin{equation}
 A=\begin{bmatrix} \label{eq:10}
-\frac{1}{T_w} & 0 & 0 & 0\\[0.5 cm]
-\frac{2K_{oil}D_p\omega_p}{V_c} & -\frac{2K_{oil}k_{leak}}{V_c} & \frac{2K_{oil}D_mk}{rV_c} & 0\\[0.5 cm]
0 & -\frac{rD_mk\eta_m}{J_w + mr^2} & -\frac{b}{J_w + mr^2} & 0\\[0.5 cm]
0 & 0 & 1 & 0
\end{bmatrix}
\end{equation}

where $k_{leak} = k_{1,p} + k_{1,m}$, the input matrix by
\begin{equation} \label{eq:11}
B = \begin{bmatrix}
\frac{1}{T_w} & 0 & 0 & 0
\end{bmatrix}^{T}
\end{equation}
the output matrix by
\begin{equation} \label{eq:12}
C = \begin{bmatrix}
0 & 0 & 0 & 1
\end{bmatrix}
\end{equation}
and disturbance by 
\begin{equation} \label{eq:13}
  d = \frac{r^2mg}{J_w + mr^2} + \widetilde{d} 
\end{equation}
where $\widetilde{d}$ is the disturbance caused due to unmodelled dynamics, nonlinear friction, vibrations and parameter uncertainty. The parameters and their definitions for the hydraulic drive of the winch are shown in Table~\ref{tb:winch_data}. Since the limits in pressure relief valve are not usually reached in practise, the oil supply does not significantly affect the governing dynamics of the winch and hence the leakage of oil is not modeled in this paper.

A more comprehensive explanation of this model can be found in \cite{zinage2020comparative}. It is assumed that the cable does not lose tension throughout the operation and the net heave time history of the vessel is already known from predictions prior to the implementation of RL based control. 

\begin{table}[htbp]
\begin{center}
\captionsetup{width=8.5cm}
\caption{Data Values}
\label{tb:winch_data}
\resizebox{8.5 cm}{!}{\begin{tabular}{lcr}
\textbf{Parameter Name} & \textbf{Parameters} & \textbf{Value}\\\hline
Acceleration due to gravity & $g$  & $9.8$  $m/s^2$\\
Bulk modulus of hydraulic fluid & $K_{oil}$ & $1.8\times10^9$ $N/m^2$\\
Volume of hydraulic lines & $V_c$ & $2\times10^{-3}$ $m^3$ \\
Maximum pump displacement & $D_p$ & $40\times10^{-6}$ $m^3$ \\
Fixed displacement of motor & $D_m$ & $4\times10^{-6}$ $m^3$\\
Rotation rate of pump & $\omega_p$ & $45$ $Hz$ \\
Leakage constant of pump & $k_{1,p}$ & $0$\\
Leakage constant of motor & $k_{1,m}$ & $0$\\
Time constant & $T_w$ & $1$ $s$ \\
Gear transmission ratio & $k$ & $200$ \\
Radius of winch & $r$ & $0.5$ $m$ \\
Efficiency of motor & $\eta_m$ & $0.65$ \\
Inertia of the winch & $J_w$ & $150$ $kg m^2$\\
Viscous friction of the winch & $b$ & $1\times10^{4}$ $kg m^2/s$\\
Mass of the payload & $m$ & $1000$ $kg$\\ \hline
\end{tabular}}
\end{center}
\end{table}

\section{Reinforcement learning based controller}

In RL, the agent interacts with the environment by taking actions and observing the state and the reward without any knowledge of the dynamics of the environment. The goal in RL algorithm is to find the optimal policy $\pi^*$ that selects the optimal control actions $u_{t}^*$ as shown below
\begin{equation} \label{eq:14}
    u_{t}^* = \pi^*(s_t) = \argmax_{\pi} Q^{\pi}(s_t,u_t)
\end{equation}
that maximises the Q value, which is the expected value of the total discounted reward when an action $u_t$ is taken in state $s_t$. Mathematically, it can be defined as 
\begin{equation} \label{eq:15}
    Q^{\pi}(s_t,u_t) =  \mathbb{E}_{\pi}[R_t|s_t, u_t]
\end{equation}
where the total discounted reward $R_t$ is given by
\begin{equation} \label{eq:16}
    R_t = \sum_{i=0}^{\infty} \gamma^{i}r_{t+i+1}
\end{equation}
Here, $r$ is the reward obtained at each time step and $\gamma$ is the discount factor. The discount factor $\gamma$ determines how much the agent values the reward at the current time step as compared to rewards obtained in the future. The optimal policy $\pi^*$ is found by using the previous history of the states visited by the agent and the rewards collected by it during its interaction with the environment. By this the agent generates experience which is then used to improve the policy. The action-value function written in a recursive format (also known as the Bellman equation) is given by
\begin{equation} \label{eq:17}
    \resizebox{.85\hsize}{!}{$Q^{\pi}(s_t,u_t) = \mathbb{E}_{r_t,s_{t+1} \sim \mathcal{E}}[r(s_t,u_t) + \gamma \mathbb{E}_{u_{t+1}\sim\pi}[Q^\pi(s_{t+1},u_{t+1})]]$}
\end{equation}
where the state $s_{t+1}$ is observed from environment $\mathcal{E}$ due to an action $u_t$ selected from state $s_t$. It is further assumed that the action $u_{t+1}$ is also selected according to the policy $\pi$. Since DDPG uses a deterministic policy and the state transition is deterministic under a selected action in this problem, the above equation then becomes
\begin{equation} \label{eq:18}
    Q^\mu(s_t,u_t) = r(s_t,u_t) + \gamma Q^\mu(s_{t+1},\mu(s_{t+1}))
\end{equation}
where $\mu$ represents the deterministic policy function.

DDPG adopts an actor-critic framework where both the policy and action-value functions are learnt using neural networks. The actor network takes in the input the current state of the agent and provides the action to be taken according to the policy as its output. The critic network takes in the action and the state as the inputs and provides the Q value as the output. The goal of the critic network is to minimise the mean square temporal difference (TD) error:
\begin{equation} \label{eq:19}
L = \frac{1}{N}\sum_{i=1}^{N}(Q(s_i,u_i|\theta^{Q})-y_i)^2
\end{equation}
where $\theta^Q$ represents the parameters of the critic network, $N$ is the sample batch size, and $y_i$ is the temporal difference (TD) target given by 
\begin{equation} \label{eq:20}
    y_i = r(s_i,u_i) + \gamma Q(s_{i+1},\mu(s_{i+1}))
\end{equation}
The TD error is defined as difference between the evaluated Q value and the TD target $y_i$. If $y_i$ is calculated using the same network used by the critic, it may be very hard to converge. So a target critic $Q'(s,u|\theta^{Q'})$ and target actor $\mu'(s|\theta^{\mu'})$ is introduced. The parameters $\theta^{Q'}$ and $\theta^{\mu'}$ is updated using a soft method given by 
\begin{equation} \label{eq:21}
\begin{split}
    \theta^{Q'} \leftarrow \tau \theta^{Q} + (1 - \tau)\theta^{Q'}\\
    \theta^{\mu'} \leftarrow \tau \theta^{\mu} +(1 - \tau)\theta^{\mu'}
\end{split}
\end{equation} 
where $\tau\ll1$. So, the TD target $y_i$ can be expressed as 
\begin{equation} \label{eq:22}
    y_i = r(s_i,u_i) + \gamma Q'(s_{i+1},\mu'(s_{i+1}|\theta^{\mu'})|\theta^{Q'})
\end{equation}

The aim of the actor network is to maximise the expected accumulated reward $J$ whose gradient is given by
\begin{equation} \label{eq:23}
    \resizebox{.85\hsize}{!}{$\nabla_{\theta^{\mu}}J \approx \frac{1}{N}\sum_{i=1}^{N}\nabla_{u}Q(s,u|\theta^{Q})|_{s_i,u = \mu(s_i)}\nabla_{\theta^{\mu}}\mu(s|\theta^{\mu})|_{s_{i}}$}
\end{equation}

where $s_i$ is sampled from the replay buffer $\mathcal{D}$. 
\begin{algorithm}
\caption{DDPG algorithm} 
\begin{algorithmic}[1]
    \State Randomly initialise critic network $Q(s,u|\theta^Q)$ and actor $\mu(s|\theta^\mu)$ with weights $\theta^Q$ and $\theta^\mu$
    \State Set target parameters equal to main parameters $\theta^{Q'}\leftarrow\theta^{Q}$, $\theta^{\mu'}\leftarrow\theta^\mu$
    \State Empty replay buffer $\mathcal{D}$
    \For {$episode = 1,M$}
    \State Initialise a random noise $\mathcal{N}$ for action exploration
    \State Receive initial observation state $s_{0}$
    \For {t = 1,T}
    \State Select action $$u_{t} = \text{clip}(\mu(s_t|\theta^\mu)+\mathcal{N}_t,a_\text{low},a_\text{high})$$
    \State Execute action $u_t$ in the environment $\mathcal{E}$
    \State Observe new state $s_{t+1}$ and reward $r_t$  
    \State Store $(s_t,a_t,r_t,s_{t+1})$ in $\mathcal{D}$
    \State Sample a random minibatch of $N$ transitions $(s_i,a_i,r_i,s_{i+1})$ from $\mathcal{D}$
    \State Compute target $$y_i = r_i + \gamma Q'(s_{i+1},\mu'(s_{i+1}|\theta^{\mu'})|\theta^{Q'})$$
    \State Update the critic network by minimising the loss
    \[L = \frac{1}{N}\sum_{i=1}^{N}(Q(s_i,u_i|\theta^{Q})-y_i)^2\]
    \State Update the actor policy by applying the following gradient:
    \[\nabla_{\theta^{\mu}}J \approx \frac{1}{N}\sum_{i=1}^{N}\nabla_{u}Q(s,u|\theta^{Q})|_{s_i,u = \mu(s_i)}\nabla_{\theta^{\mu}}\mu(s|\theta^{\mu})|_{s_{i}}\]
    \State Update the target networks with
     \[\theta^{Q'} \leftarrow \tau \theta^{Q} + (1 - \tau)\theta^{Q'}\]
     \[\theta^{\mu'} \leftarrow \tau \theta^{\mu} +(1 - \tau)\theta^{\mu'}\]
    \EndFor
    \EndFor
\end{algorithmic} 
\end{algorithm}

\begin{figure}[ht]
\begin{center}
\includegraphics[width=8.4cm]{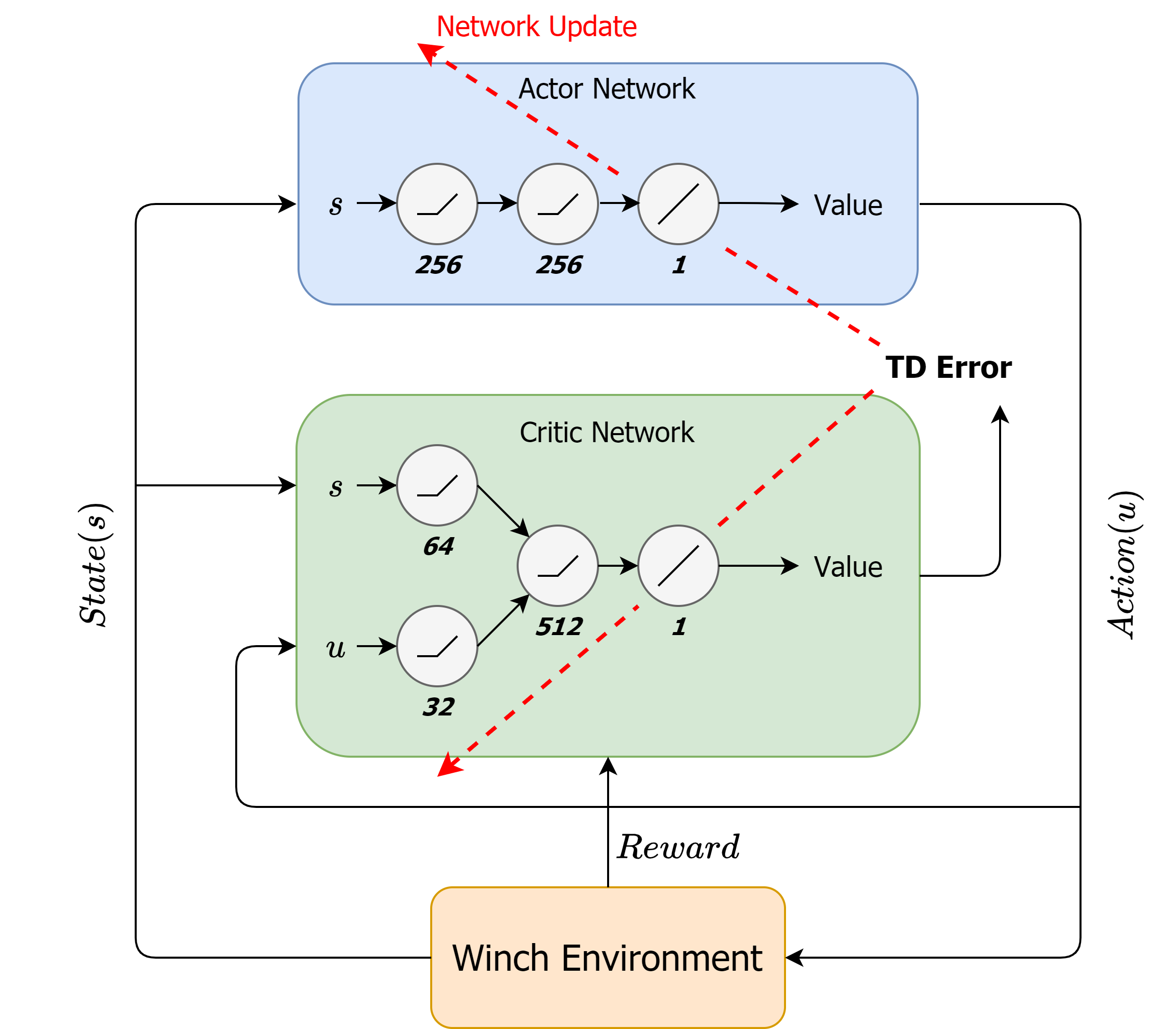}
\caption{DDPG Architecture} 
\label{fig:policy}
\end{center}
\end{figure}

The RL agent is programmed using Keras with a Tensorflow backend. The training has been performed on Nvidia GeForce GTX 1060 16GB GPU. The state space $s \in \mathbb{S}$ chosen for the application of RL is defined as
\begin{equation} \label{eq:24}
    \mathbb{S} = \{z_w,\dot{z}_w,z_\text{winch},\dot{z}_\text{winch}\}
\end{equation}
where $z_w$ is the length of the reeled rope and $z_\text{winch}$ is the net heave at the winch due to the motion of the vessel in waves. The action space $u \in \mathbb{A}$ is defined as
\begin{equation} \label{eq:25}
    \mathbb{A} = \{u_p\}
\end{equation}
where $u_p$ is the control input provided to the winch model.

Fig.~\ref{fig:policy} shows a schematic diagram of the structure of the DDPG architecture used. The pseudo code of the DDPG algorithm is shown above. The actor network is composed of two fully connected layers with 256 neurons each whereas the critic network is composed of a fully connected layers with 64 and 32 neurons for inputs $s$ and $u$ respectively followed by a fully connected layer of 512 neurons as shown in Fig.~\ref{fig:policy}. A rectified linear unit function is used as the activation function for each neuron. 

Since the goal in active heave compensation is to minimise the compensated motion without producing any chattering of the payload motion, the reward $r$ is defined as follows
\begin{equation} \label{eq:26}
  r=
        \begin{cases}
                                   1 - 20 e_z - \dot{e}_z & \text{$e_z \leq 0.05$ m} \\
                                   - 10 e_z - 2 \dot{e}_z & \text{$e_z > 0.05$ m}
        \end{cases}
\end{equation}

where \(e_z = \abs{z_w + z_\text{winch}}\) and \(\dot{e}_z = \abs{\dot{z}_w +\dot{z}_\text{winch}}\) respectively.

\begin{figure}[ht]
\begin{center}
\includegraphics[width = 8.4cm]{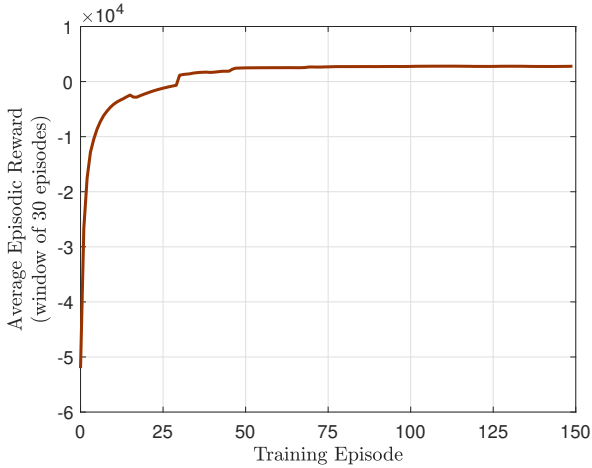}
\caption{Learning curve of the RL agent} 
\label{fig:reward}
\end{center}
\end{figure}

The RL based controller is trained for $150$ episodes for $3000$ training steps with a sample time of $0.1$ sec. The reference used for training is a 300 sec net heave time history at the winch in moderate sea state. During the simulation, the initial conditions of the states are sampled from a uniform distribution. The range of state 1 is $(-1,1)$, the range of state 2 is $(-10^6,10^6)$, the range of state 3 is $(-0.1,0.1)$, and the range of state 4 is $(0,1)$. The sample batch size is set to $128$ with a replay buffer capacity of $50000$. For training the network, an Adam optimiser is used for both the actor and the critic network. The learning rate of the actor and the critic network are set to $0.001$ and $0.002$ respectively. The target network transition gain $\tau$ is set as $0.005$ and the discount factor $\gamma$ is set as $0.998$. For better exploration, an Ornstein–Uhlenbeck action noise with $\theta = 0.15$, $\mu = 0$, and $\sigma = 0.0005$ is used. Fig~\ref{fig:reward} shows the learning curve achieved as a result of the training. The average episodic reward in Fig~\ref{fig:reward} indicates the average of the rewards received in the last 30 episodes.

For comparison, an PD controller is used, with the tuned gains as shown below
\begin{equation} \label{eq:27}
  K_p = 5.86,~ K_d = 5.46,~ T_f = 0.03 
\end{equation}
where $K_p$, $K_d$ are the controller gains and $T_f$ is the time constant for noise filter. The gains are tuned using loop shaping approach (\cite{zinage2020comparative}). The control input for this control law in the Laplace domain is given by 
\begin{equation} \label{eq:28}
    U(s) = \left (K_{p} + \frac{K_{d}s}{1 + T_{f}s} \right ) E(s)
\end{equation}
where $U(s)$ and $E(s)$ are the Laplace transform of $u_{p}(t)$ and \(e(t) \equiv -(z_w(t) + z_\text{winch}(t))\) respectively.

\section{Simulation Results}

The following four cases are analysed to understand the advantages and limitations of using RL based control over classical control.

\subsection{Heave compensation with no disturbance and no noise}

In this study, heave compensation is defined as the ratio of the difference between the RMS value of uncompensated and compensated net heave at the winch to the RMS value of uncompensated net
heave at the winch. Table~\ref{tb:heave_compensation} shows the comparison of the compensation performance between RL and PD controller for different sea states.

\begin{table}[htbp]
\begin{center}
\captionsetup{width=8.5cm}
\caption{Heave compensation in different sea states}
\label{tb:heave_compensation}
\resizebox{8.5 cm}{!}{\begin{tabular}{ccc}
\textbf{Sea State} & \textbf{RL-Control} & \textbf{PD-Control}\\\hline
Slight ($H_s$ = $1.5$ m, $T_p$ = $6$ s) & $96.9 \%$  & $86.6 \%$\\
Moderate ($H_s$ = $4$ m, $T_p$ = $9$ s) & $99.3 \%$ & $89.8 \%$\\
Rough ($H_s$ = $6$ m, $T_p$ = $12$ s) & $98.1 \%$ & $92.1 \%$ \\
Very Rough ($H_s$ = $8.5$ m, $T_p$ = $14$ s) & $95.53 \%$ & $83.26 \%$ \\ \hline
\end{tabular}}
\end{center}
\end{table}

\begin{figure}[ht]
\begin{center}
\includegraphics[width=8.4cm]{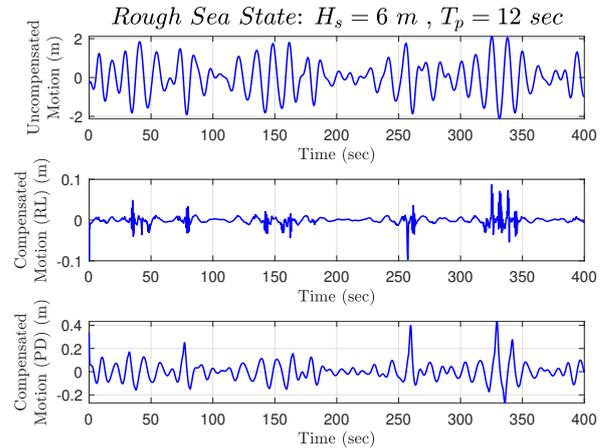}
\caption{Uncompensated and compensated motion time histories at the winch for rough sea state.} 
\label{fig:case_1}
\end{center}
\end{figure}

It can be seen that the RL based control is able to demonstrate a better heave compensation performance than the PD control for all the four sea states. As the sea state keeps on increasing the effect of saturation in the swash angle is observed more often thereby leading to a decreased compensation performance in higher sea states. However, as per Table~\ref{tb:heave_compensation} it can be seen that the compensation performance of the RL controller almost remained constant irrespective of sea state. This is indicative that the RL based control is able to handle the saturation in the swash angle better than the PD control. Fig.~\ref{fig:case_1} shows the plot of the uncompensated motion and the compensated motion time histories in rough sea state.

\subsection{Heave compensation with an offset}

\begin{figure}[H]
\begin{center}
\includegraphics[width=8.4cm]{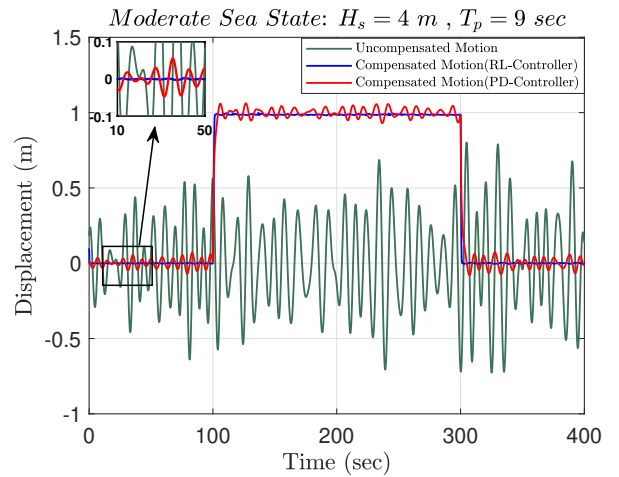}
\caption{Ability of the controllers to track the offset}
\label{fig:case_2}
\end{center}
\end{figure}

Fig.~\ref{fig:case_2} shows the ability of the both the controllers in tracking an offset for a period of $200$ secs for a wave incident angle of $135$ degrees. It can be seen that the RL controller performed better than the PD controller while tracking an offset. This type of offset tracking is particularly important in many offshore operations where a load is lowered from an offshore crane onto a floating structure. Topside installation of offshore platforms is an example where offset tracking is an important requirement.

\subsection{Heave compensation with disturbance but no noise}

\begin{figure}[ht]
\begin{center}
\includegraphics[width=8.4cm]{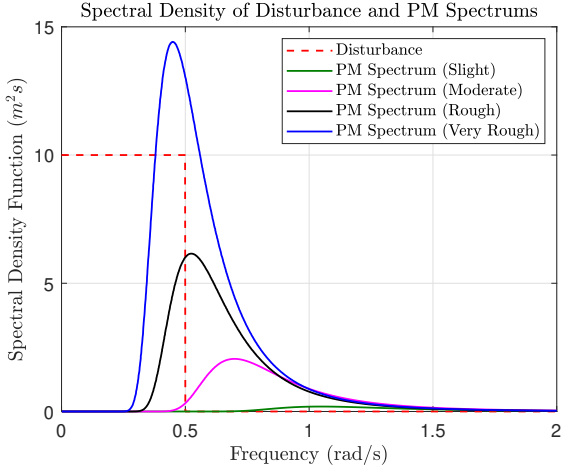}
\caption{Spectrum of disturbance and PM spectra's for different sea states} 
\label{fig:case_3_1}
\end{center}
\end{figure}

In order to analyze the ability of the controllers to reject disturbance $d$, a disturbance time history is generated with a constant spectrum of spectral density value $10$ with a cut in and cut off frequencies of $0$ $rad/s$ and $0.5$ $rad/s$ respectively. This disturbance $d$ can be caused due to constant payload torque, unmodelled dynamics, nonlinear friction, vibrations and parametric uncertainty in the real physical system. Fig.~\ref{fig:case_3_1} shows the plot of the spectrum for the disturbance and the wave elevations in different sea states. In this study, it is assumed that the disturbance entered the system only through the third state (i.e reeled

\begin{figure}[ht]
\begin{center}
\includegraphics[width=8.4cm]{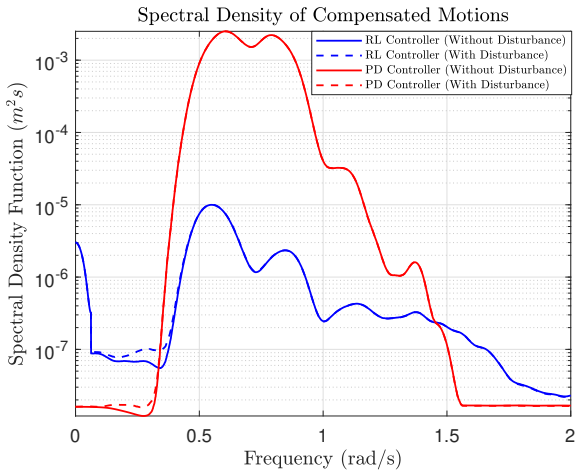}
\caption{Spectrum of compensated motions in moderate sea state} 
\label{fig:case_3}
\end{center}
\end{figure}

rope velocity $\dot{z}_w$). Since the ability of both the control strategies in rejecting disturbances is independent of sea state, the simulation results is analyzed only for the moderate sea state. Fig.~\ref{fig:case_3} shows the power spectral density of the compensated motions for both the controllers in the presence and absence of disturbance. It can be observed that both controllers are good at disturbance rejection.

\subsection{Heave compensation with noise but no disturbance}

The following two cases are analyzed in this section to test the ability of the controllers in attenuating measurement noise: one with a low noise having spectral density value of $10^{-6}$ with a cut in and cut off frequency of $3.14$ $rad/s$ and $30$ $rad/s$ respectively and other with a high noise having spectral density value of $10^{-3}$ with the same cut in and cut off frequencies. In order to replicate an actual measured output signal, the noise generated is added to $z_{w}$ and $\dot{z}_{w}$ and provided as an input to both the controllers. 

The signal to noise ratio (SNR) is usually defined in decibels as
\begin{equation} \label{eq:29}
    \text{SNR} = 10 \log_{10} \left( \frac{\sigma_{signal}^2}{\sigma_{noise}^2} \right) = 20 \log_{10} \left( \frac{\sigma_{signal}}{\sigma_{noise}} \right)
\end{equation}
In this study, $\sigma_{signal}$ is taken as the RMS value of the uncompensated net heave motion at the winch and $\sigma_{noise}$ is taken as the RMS value of the noise. A constant noise spectral density of $10^{-6}$ between $3.14$ $rad/s$ and $30$ $rad/s$ corresponds to a SNR value of $34.53$ $dB$ whereas a constant noise spectral density of $10^{-3}$ with the same cut in and 

\begin{figure}[ht]
\begin{center}
\includegraphics[width=8.4cm]{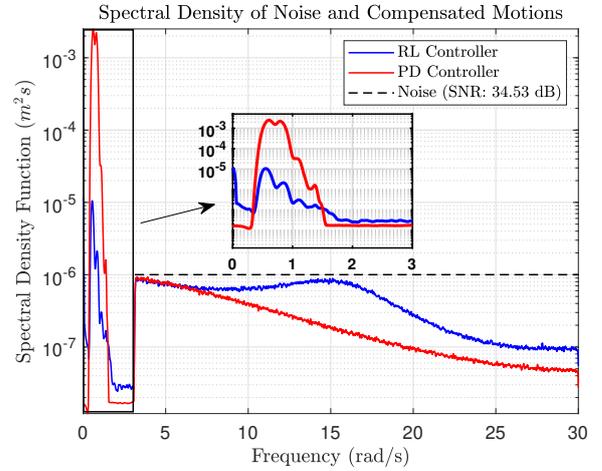}
\caption{Spectrum of compensated motions for low noise in moderate sea state} 
\label{fig:case_4_1}
\end{center}
\end{figure}

cut off frequencies corresponds to a SNR value of $4.56$ $dB$ for moderate sea state.

Fig.~\ref{fig:case_4_1} shows the power spectral density of the noise and compensated motions for both the strategies when a low noise is included. From Fig.~\ref{fig:case_4_1}, it can be seen that in high SNR case, the PD controller is able to perform better in attenuating the noise at higher frequencies. Also, in this case there is no effect of noise seen on the compensation ability of both the controllers as per Fig.~\ref{fig:case_3} and zoomed plot inside Fig.~\ref{fig:case_4_1}.

\begin{figure}[ht]
\begin{center}
\includegraphics[width=8.4cm]{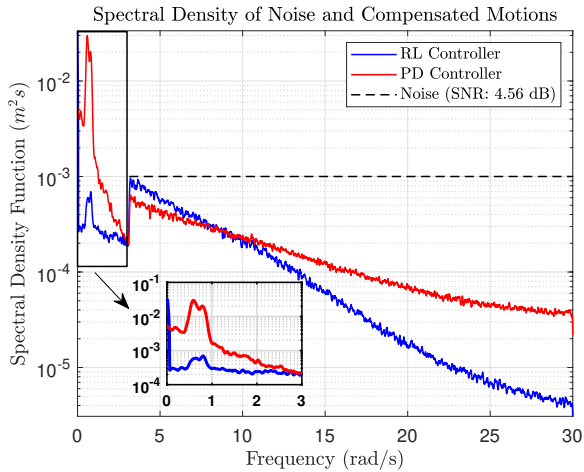}
\caption{Spectrum of compensated motions for high noise in moderate sea state} 
\label{fig:case_4_2}
\end{center}
\end{figure}

Fig.~\ref{fig:case_4_2} shows the power spectral density of the noise and compensated motions for each of the control strategies when a high noise is included. As per Fig.~\ref{fig:case_4_2}, it can be seen that RL controller performed better in attenuating higher magnitude noise. Saturation in the normalized swash angle is observed for both the controllers in this case, and due to this the compensation ability is significantly effected. As per Fig.~\ref{fig:case_3} and zoomed plot inside Fig.~\ref{fig:case_4_2}, the PD controller is found to be more affected due to the noise as compared to the RL based controller. When it came to high noise attenuation the RL based controller performed better than the PD controller as shown in Fig.~\ref{fig:case_4_2}.

\section{Conclusion}

In this paper, a deep reinforcement learning based controller is proposed for active heave compensation (AHC). A deep deterministic policy gradient (DDPG) algorithm is used to develop a controller for AHC. The control policy is trained by simulating the winch environment for several episodes in a moderate sea state. The extracted control policy is then tested on four different sea states to validate its heave compensation ability. A Proportional-Derivative (PD) controller is tuned and used to compare the results of the RL based controller. It is found that the RL based controller is able to provide a better compensation performance than the PD controller in all four sea states. Both the controllers are reasonably good in disturbance rejection. The PD controller is able to attenuate noise better in a low noise scenario. However, in a high noise scenario, the RL based controller is able to provide better noise attenuation when measurement noise is added to the feedback signal. The RL based controller is also able to handle the saturation dynamics better than the PD controller.

\bibliography{main}         
                                                   
\end{document}